\documentclass{iopconfser}

\usepackage{amsmath}
\usepackage{tikz}

\usetikzlibrary{arrows, positioning,arrows.meta}
\usetikzlibrary{backgrounds}
\usetikzlibrary{calc}

\usepackage{svg}
\usepackage{array} 

\usepackage{graphicx}
\usepackage{newtxtext}
\usepackage{newtxmath}
\usepackage[numbers]{natbib} 
\usepackage{hyperref}

\usepackage{enumitem}

\usepackage{bm}
\usepackage[caption=false,justification=raggedleft]{subfig}
\usepackage{algorithm}
\usepackage{algpseudocode}

\usepackage{placeins}

\usepackage[abs]{overpic}  

\begin{document}

\title{Towards extreme event prediction of  turbulent flows with quantized local reduced-order models.}

\author{Antonio Colanera$^{1,2}$ and Luca Magri$^{1,2}$}

\affil{$^1$Department of Mechanical and Aerospace Engineering, Politecnico di Torino, Turin, Italy}
\affil{$^2$Department of Aeronautics, Imperial College London, London, United Kingdom}


\email{l.magri@imperial.ac.uk}

\begin{abstract}
This work develops quantized local reduced-order models (ql-ROMs) of the turbulent Minimal Flow Unit (MFU) for the analysis and interpretation of intermittent dissipative dynamics and extreme events. The ql-ROM combines data-driven clustering of the flow state space with intrusive Galerkin projection on locally defined Proper Orthogonal Decomposition (POD) bases. This construction enables an accurate and stable low-dimensional representation of nonlinear flow dynamics whilst preserving the structure of the governing equations. 
The model is trained on direct numerical simulation data of the MFU. When deployed, the ql-ROM is numerically stable for long-term integration, and correctly infers the statistical behavior of the kinetic energy and dissipation observed of the full-order system. A local modal energy-budget formulation is employed to quantify intermodal energy transfer and viscous dissipation within each region of the attractor. The analysis reveals that dissipation bursts correspond to localized energy transfer from streamwise streaks and travelling-wave modes toward highly dissipative vortical structures, consistent with the self-sustaining process of near-wall turbulence.
Beyond reduced modeling, the ql-ROM framework provides a pathway for the reduced-space characterization and potential prediction of extreme events. ql-ROM offer an interpretable and computationally efficient framework for the analysis and prediction of extreme events in turbulent flows.

\end{abstract}

\section{Introduction}
Reduced-order models (ROMs) provide low-dimensional surrogates of high-dimensional fluid systems by projecting the governing equations of the full-order model (FOM) onto a small set of dominant modes (intrusive ROMs). 
Classical approaches based on Proper Orthogonal Decomposition (POD) and Galerkin projection~\citep{rowley2017arfm, Berkooz1993, Holmes1996} have successfully captured coherent structures in transitional and turbulent flows, offering physically interpretable and computationally efficient representations. However, a single ROM of the entire flow field---also referred to as a ``global ROM"---often suffers from stability and accuracy issues in nonlinear and multiscale regimes, because the assumption of a single subspace is inadequate to describe the complex geometry of turbulent attractors \cite{Floryan2022}. To overcome these limitations, several extensions have been developed, including closure modeling~\citep{Duraisamy2019, Xiao2019}, adaptive or localized bases~\citep{Amsallem2008, Peherstorfer2014}, and data-driven or hybrid methods~\citep{Brunton2020, Benner2015}. Among these, cluster-based and local ROMs have proven particularly effective in turbulent flow applications~\citep{Kaiser2014, Nair2019jfm,li2020jfm, Colanera2024a,Colanera2024b}. 
By partitioning the state space into regions of similar flow behavior and constructing local models around representative states, these approaches can better capture nonlinearity and intermittency. 
The recently introduced quantized local reduced-order model (ql-ROM)~\citep{Colanera2025b} proposes this paradigm by combining clustering-based manifold quantization with intrusive Galerkin projection on local bases centered around cluster centroids. 

Among the various configurations used to study reduced representations of turbulence, the Minimal Flow Unit (MFU) provides the smallest computational domain capable of sustaining near-wall turbulence whilst preserving the essential mechanisms of the self-sustaining process (SSP). Originally introduced by Jiménez and Moin~\citep{JimenezMoin1991}, the MFU allows the isolation and analysis of the fundamental dynamical interactions between streamwise streaks, quasi-streamwise vortices, and travelling-wave modes. Hamilton, Kim, and Waleffe~\citep{Hamilton1995} demonstrated that the regeneration of near-wall turbulence is governed by a cyclic sequence of streak formation via the lift-up mechanism, streak instability and breakdown into travelling-wave perturbations, and subsequent vortex regeneration. 
This conceptual framework was later formalized by Waleffe~\citep{Waleffe1997}, who developed a reduced theoretical model describing the nonlinear coupling among rolls, streaks, and waves responsible for maintaining turbulence. 
Subsequent studies identified the connection between streak breakdown events and dissipation bursts within the MFU~\citep{KAWAHARA2001, SCHOPPA2002}, revealing that these bursts mark the transition between the regeneration and decay phases of the SSP. 
Further reviews and analyses~\citep{Jimenez2013} reinforced the interpretation of the MFU as a minimal yet dynamically complete system that encapsulates the essential nonlinear energy-transfer processes governing wall-bounded turbulence.

A relevant feature of wall-bounded turbulence is the occurrence of rare but dynamically significant extreme events, sudden bursts of dissipation or kinetic energy production that have a major impact on drag, noise, and flow control. 
Understanding their underlying mechanisms and developing predictive indicators is of both scientific and practical importance. 
Itano and Toh~\citep{Itano2001} described the bursting process in wall turbulence as a cyclic regeneration of near-wall streaks and vortices, while Hack and Schmidt~\citep{Hack2020} characterized extreme dissipation events in wall turbulence and their spatial organization. 
Blonigan, Farazmand, and Sapsis~\citep{Blonigan2019} analyzed the predictability of such extreme events using finite-time Lyapunov vector alignment as an early-warning indicator, and Ciola \emph{et al.}~\citep{Ciola2023} computed nonlinear optimal perturbations of turbulent channel flow as robust precursors of high-dissipation events. 
At high-pressure transcritical conditions, El~Mansy \emph{et al.}~\citep{ElMansy2024} demonstrated that MFU configurations remain dynamically relevant, while Yin, Hwang, and Vassilicos~\citep{Yin2024} recently elucidated the dynamics of turbulent energy and dissipation across scales in wall-bounded turbulence. 
Recent advances in data-driven modeling have also demonstrated the feasibility of learning precursors and control strategies for extreme events directly from high-dimensional chaotic dynamics~\citep{Racca2021,Doan2021,Racca2022,Oezalp2025}.
Within this context, reduced-order modeling offers a promising framework for detecting and analyzing such intermittent phenomena. 
In particular, cluster-based ROMs such as the ql-ROM enable the identification of phase-space regions associated with extreme behavior, provide localized quantification of modal energy transfers and dissipation, and offer a low-cost yet interpretable tool for the analysis and potential prediction of extreme events in turbulent flows.  

This work aims to (i) construct a ql-ROM for the turbulent MFU that remains stable over long integrations and accurately reproduces key statistics; (ii) introduce a local modal energy-budget analysis to quantify intermodal transfers and viscous dissipation within clusters to characterize dissipation bursts.

The paper is organized as follows. Section~\ref{sec:methods} presents the methodology, including the MFU setup, the Galerkin POD formulation, the ql-ROM framework, and the local modal energy budget approach. Section~\ref{sec:results} discusses the results, highlighting the performance of the ql-ROM and its ability to characterize dissipation bursts. Conclusions are summarized in Section~\ref{sec:conclusions}.

\section{Methodology}\label{sec:methods}
In this section, the methodologies employed in this work are presented. 
Section~\ref{sec:methodology_mfu} explains the Minimal Flow Unit (MFU) configuration and the numerical setup. 
Section~\ref{sec:methodology_mfu:pod} outlines the construction of the Galerkin POD reduced-order model. 
The quantized local ROM (ql-ROM) framework, which combines clustering and local projection, is detailed in Section~\ref{sec:methodology_mfu:qlrom}. 
Finally, Section~\ref{sec:methodology_mfu:budget} describes the local modal energy budget approach used to analyze intermodal energy transfers and dissipation mechanisms.

\subsection{Minimal Flow Unit}
\label{sec:methodology_mfu}

\begin{figure}[t]
        \centering
        \vspace{0.6cm} 
        \includegraphics[decodearray={1 0.  1 0  1 0},width=0.7\textwidth]{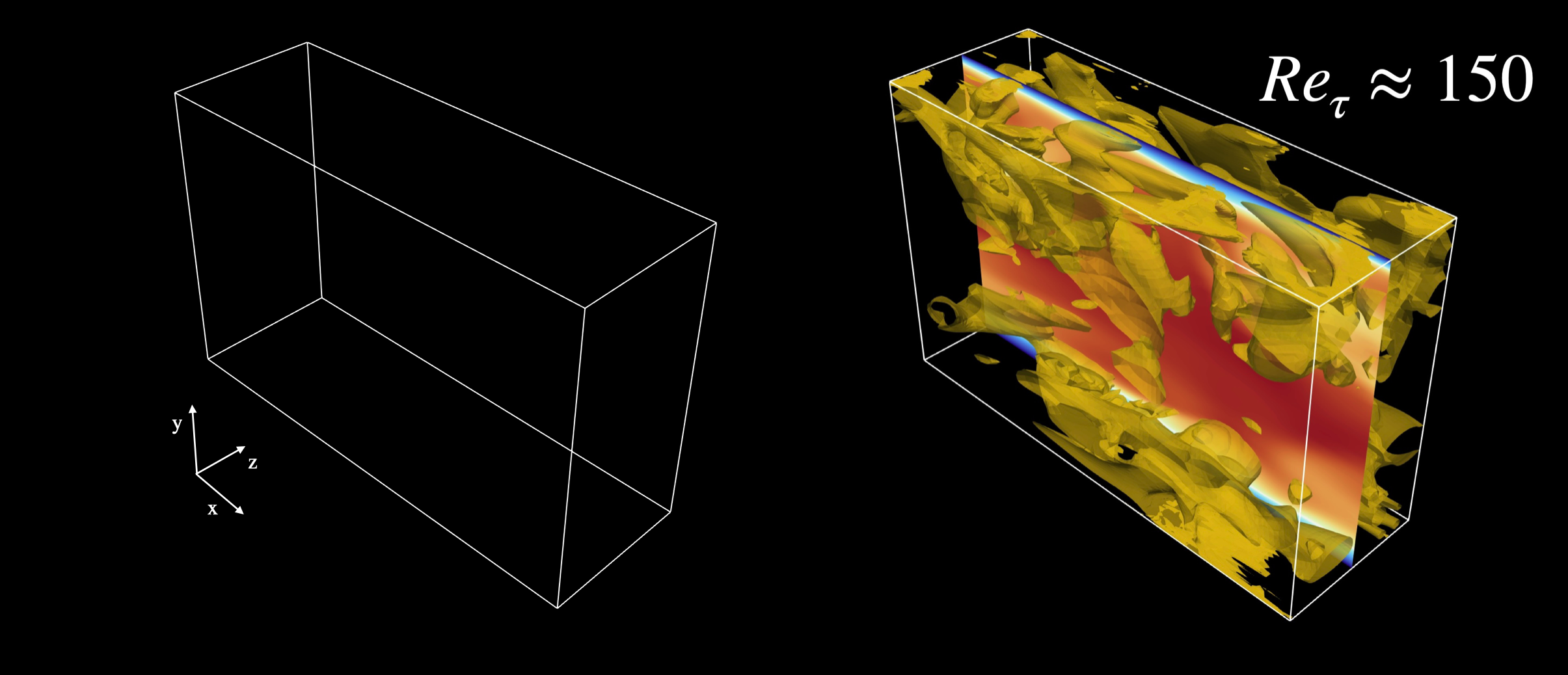}
  \caption{Minimal Flow Unit. Computational setup. The left panel shows the computational domain. The right panel displays a representative snapshot of the simulated flow field, including the iso-surface of the $Q$-criterion ($Q = 0.05$) and the axial velocity distribution on the mid-plane $z=0$.} 
    \label{fig:MFUsketch}
\end{figure}

\begin{figure}
        \centering
        \vspace{0.6cm} 
        \includegraphics[decodearray={1 0.  1 0  1 0},width=0.99\textwidth]{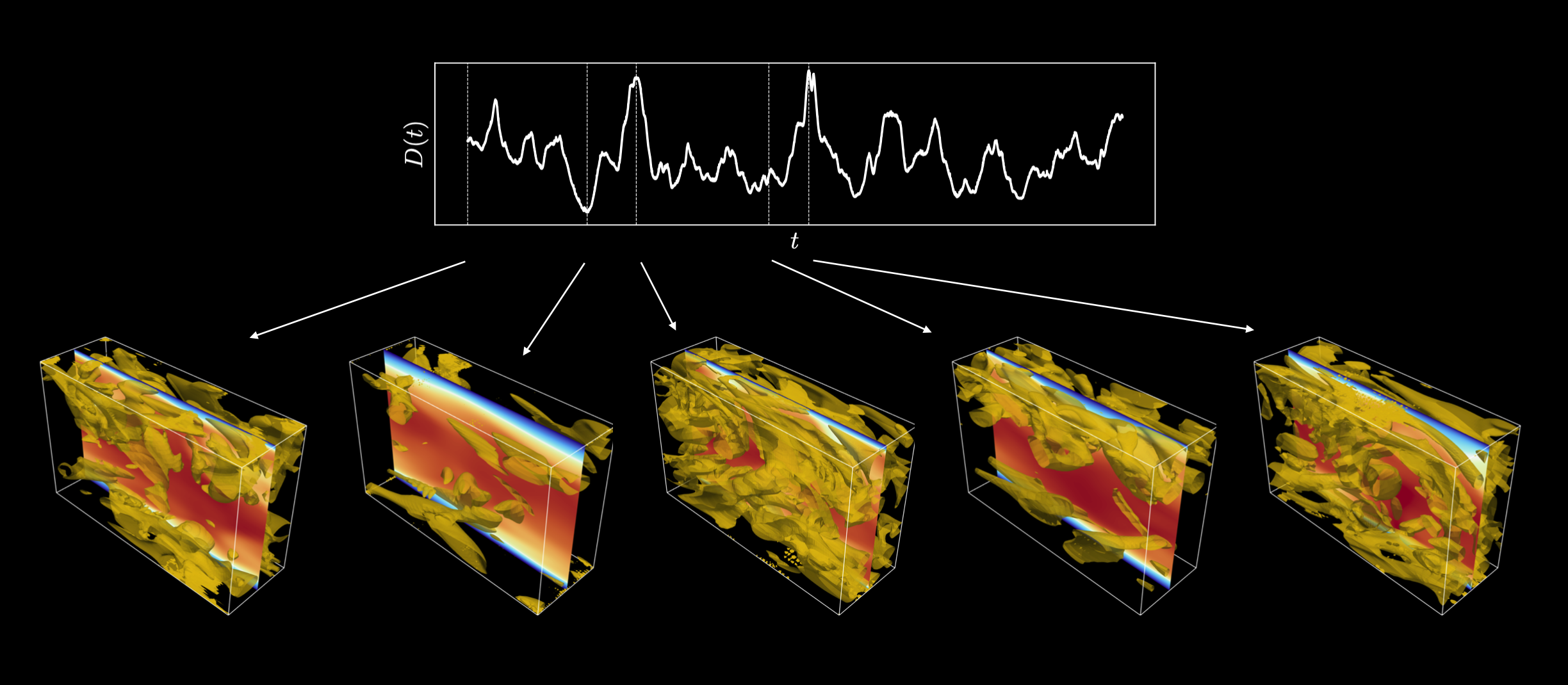}
  \caption{Intermittent dynamics of the MFU. 
    The upper panel shows the temporal evolution of the domain-averaged dissipation $D(t)$, 
    highlighting intense dissipative bursts (extreme events). 
    Dashed lines mark high-dissipation peaks and pre-burst states, illustrated below, 
    where partial relaminarization is followed by rapid regeneration and energy transfer leading to dissipation bursts.
   } 
    \label{fig:mfuinterm}
\end{figure}

We consider an incompressible flow in a minimal flow unit (MFU) with periodic boundary
conditions in the homogeneous directions and no-slip condition on walls. The velocity field $\mathbf{u}(\mathbf{x},t)$ evolves in the domain $\Omega$ under an external
forcing $\mathbf{f}$ (e.g., a constant pressure-gradient term), and satisfies
\begin{equation}
\frac{\partial \mathbf{u}}{\partial t} + (\mathbf{u}\!\cdot\!\nabla)\mathbf{u}
= -\frac{1}{\rho}\nabla p + \nu \nabla^2 \mathbf{u} + \mathbf{f}, 
\qquad \nabla\!\cdot\!\mathbf{u} = 0,
\label{eq:nse_mfu}
\end{equation}
where $\rho$ is the density, $\nu$ is the viscosity and $p$ is the pressure.
The MFU domain is defined as $x \in [0,\, \pi h]$, $y \in [0,\, 2h]$, $z \in [0,\, 0.34\,\pi h],$ where $h$ is the channel half-height \cite{JimenezMoin1991,Blonigan2019}. The streamwise ($x$) and spanwise ($z$) directions are periodic, and the wall-normal direction ($y$) satisfies no-slip boundary conditions. The numerical grid consists of $[96,\,129,\,32]$  discretization points along the $[x,\,y,\,z]$ directions. 
The mesh is uniform in $x$ and $z$, and exponentially refined toward the walls in the $y$ direction. The Reynolds number based on the bulk velocity $U_b$, $Re = U_b h / \nu = 2300$, corresponds to a friction Reynolds number $Re_\tau \approx u_\tau h / \nu \approx 150$, with $u_\tau$ being the friction velocity. The forcing term $\mathbf{f}$ represents the constant mean pressure gradient, or equivalently a constant bulk flux, ensuring a statistically stationary turbulent regime. Figure \ref{fig:MFUsketch} shows the computational domain and a snapshot of the flow field. 
For this set of parameters, the dynamics is intermittent \cite{JimenezMoin1991}, characterized by intense dissipative bursts, as shown in the upper panel of Figure~\ref{fig:mfuinterm}, where the temporal behaviour of the domain averaged dissipation $D$ is defined as   
\begin{equation}\label{eq:dissipU}
   D(t) = \frac{\nu}{|\Omega|} \int_{\Omega} |\nabla \times \mathbf{u}(t)|^2\, d\Omega.
\end{equation}
The dashed vertical lines in Figure~\ref{fig:mfuinterm} highlight high-dissipation peaks and snapshots taken shortly before the onset of dissipation bursts. Before a dissipation spike, the flow typically exhibits a partial relaminarization in one or both halves of the channel: near-wall streaks weaken and vortical activity is reduced. These laminar-like intervals are followed by rapid regeneration events leading to strong nonlinear interactions and energy transfer, culminating in bursts of dissipation \cite{JimenezMoin1991,Hamilton1995,XiGraham2010}.

In the MFU, all volume inner products are defined as $\langle \mathbf{a},\mathbf{b}\rangle_\Omega = \int_{\Omega}
\mathbf{a}\!\cdot\!\mathbf{b}\,\mathrm{d}\Omega$, and surface integrals vanish because of the boundary
conditions. 

\subsection{Galerkin POD ROM}
\label{sec:methodology_mfu:pod}
Our methodology is based on the construction of local intrusive Galerkin projection POD-ROMs. In this section a brief introduction of this methodology is provided. Let $\overline{\mathbf{u}}$ be a temporal mean and  $\mathbf{u}'=\mathbf{u}-\overline{\mathbf{u}}$ be the  fluctuation.
From $M$ snapshots, $\{\mathbf{u}(t^m)\}_{m=1}^M$, we compute  POD modes
$\{\mathbf{\phi}_i\}_{i=1}^r$ \cite{Berkooz1993}, and express the fluid quantities on the POD modes
\begin{equation}
\mathbf{u}'(\mathbf{x},t) \approx \sum_{i=1}^r a_i(t)\,\mathbf{\phi}_i(\mathbf{x}), 
\qquad \langle \mathbf{\phi}_i,\mathbf{\phi}_j\rangle_\Omega=\delta_{ij}.
\label{eq:pod_mfu}
\end{equation}
Similarly, we expand the pressure fluctuation $p'(\mathbf{x},t)=p(\mathbf{x},t)-\overline{p}(\mathbf{x})$ as
\begin{equation}
    p'(\mathbf{x},t)\approx \sum_{m=1}^{r_p} b_m(t)\,\psi_m(\mathbf{x}),
\end{equation}
where $\{\psi_m\}_{m=1}^{r_p}$ is the POD basis for the pressure fields and $\overline{p}(\mathbf{x})$ the a reference pressure field that can be either the mean pressure or the solution of the pressure Poisson equation for the mean velocity flow.
Galerkin projection of the momentum eq. \eqref{eq:nse_mfu} onto $\mathbf{\phi}_i$ yields the ROM for the velocity that is a quadratic form
\begin{equation}
\dot{a}_i 
= f_i 
+ \sum_{j=1}^r L_{ij}\,a_j
+ \sum_{j,k=1}^r C_{ijk}\,a_j a_k
- \nu \sum_{j=1}^r B_{ij}\,a_j
+ \sum_{j=1}^{r_p} P_{ij}\,b_j
+ \tau_i,
\qquad i=1,\dots,r,
\label{eq:rom_mfu}
\end{equation}
with coefficients
\begin{align}
f_i &= \Big\langle (\overline{\mathbf{u}}\!\cdot\!\nabla)\overline{\mathbf{u}}
- \tfrac{1}{\rho}\nabla\overline{p} + \mathbf{f},\,\mathbf{\phi}_i \Big\rangle, \nonumber\\
L_{ij} &= \Big\langle (\overline{\mathbf{u}}\!\cdot\!\nabla)\mathbf{\phi}_j
+ (\mathbf{\phi}_j\!\cdot\!\nabla)\overline{\mathbf{u}},\,\mathbf{\phi}_i \Big\rangle, \qquad
C_{ijk} = \Big\langle (\mathbf{\phi}_j\!\cdot\!\nabla)\mathbf{\phi}_k,\,\mathbf{\phi}_i \Big\rangle, \nonumber\\
B_{ij} &= \Big\langle \nabla \mathbf{\phi}_j : \nabla \mathbf{\phi}_i \Big\rangle, \qquad P_{ij} = \Big\langle \tfrac{1}{\rho}\nabla \psi_j, \,\mathbf{\phi}_i  \Big\rangle. \label{eq:coeffs_mfu}
\end{align}
$\tau_i$ in eq.~\eqref{eq:rom_mfu} accounts for truncation/closure (e.g., eddy viscosity or data-driven
corrections). For MFU periodic/homogeneous directions and no-slip walls, the continuous
nonlinear term is energy-preserving, implying $\sum_{i,j,k} a_i\, C_{ijk}\, a_j a_k=0$
for divergence-free bases; deviations in practice quantify projection/discretization error.
To enforce the divergence free condition in eq.~\eqref{eq:nse_mfu} and close the problem for the coefficients $b_j$  in eq.~\eqref{eq:rom_mfu}, we project the pressure Poisson equation
\begin{equation}
    \nabla^2 p
= -\,\nabla\!\cdot\!\big[(\mathbf u\!\cdot\!\nabla)\mathbf u\big]
+ \nabla\!\cdot\!\mathbf f,
\end{equation}
onto the pressure modes $\{\psi_m\}_{m=1}^{r_p}$ obtaining a reduced algebraic system for the $b_j$
\begin{equation}
\sum_{j=1}^{r_p} K_{ij} b_j
= g_i + \sum_{j=1}^r H_{ij} a_j + \sum_{j,\ell=1}^r G_{ij\ell} a_j a_\ell + \mathcal{B}_i,
\label{eq:pressure_rom_mfu}
\end{equation}
with coefficients
\begin{align}
	K_{ij} &:= \langle \nabla\psi_j, \nabla\psi_i\rangle, \qquad
	g_i := \langle \nabla \overline p, \nabla \psi_i\rangle
	\;-\; \langle (\overline{\mathbf u}\!\cdot\!\nabla)\overline{\mathbf u}, \nabla\psi_i\rangle
	+ \langle \mathbf f, \nabla\psi_i\rangle, \\
	H_{ij} &:= -\,\Big\langle
	(\overline{\mathbf u}\!\cdot\!\nabla)\mathbf\phi_j + (\mathbf\phi_j\!\cdot\!\nabla)\overline{\mathbf u},
	\,\nabla\psi_i\Big\rangle, \qquad 
	G_{ijl} := -\,\langle (\mathbf\phi_j\!\cdot\!\nabla)\mathbf\phi_l, \nabla\psi_i\rangle, \\
	\mathcal{B}_i &:= 
	\text{Boundary conditions}
    \label{eq:coeffPre}
\end{align}
All operators in eq.~\eqref{eq:coeffs_mfu} and eq.~\eqref{eq:coeffPre} can be preassembled once the bases are fixed. In practice we have used supremizer enrichment to eliminate or stabilize the pressure coupling \cite{BallarinManzoniQuarteroniRozza2015}.

\subsection{Quantized local reduced-order modeling (ql-ROM)}
\label{sec:methodology_mfu:qlrom}

To model the complex geometry of turbulent attractors, we adopt ql-ROM in time \cite{Colanera2025b}. 
The first step of ql-ROM consists of creating the cartography of the data manifold by quantizing it into discrete patches (clusters). The snapshot set is partitioned into $K$ clusters
$\{\mathcal{C}_k\}_{k=1}^K$ via $k$-means with Euclidean distance on state vectors or reduced
coordinates \cite{steinhaus1956,MacQueen1967,Lloyd1982,Arthur2007}. Each cluster is centered around a centroid $\mathbf{c}_k$, where $k = 1, \ldots, K$, that physically represents the mean state of each cluster and is computed as
    \begin{equation}
	\mathbf{c}_k = \frac{1}{n_k} \sum_{\mathbf{u}_m \in \mathcal{C}_k} \mathbf{u}_m = \frac{1}{n_k} \sum_{m=1}^M \chi_k^m \mathbf{u}_m,
\end{equation}
where $\mathcal{C}_k$ denotes the $k$th cluster and each entry of the characteristic function $\chi_k^m$ is
	\begin{equation}
		\chi_k^m =
		\begin{cases}
			1, & \text{if } k = \beta(\mathbf{u}_m). \\
			0, & \text{otherwise}.
		\end{cases}
	\end{equation}
 The cluster affiliation function, $\beta(\mathbf{u})$, is defined as the function that assigns a  point of the phase space $\mathbf{u}$ to the index of its closest centroid
	\begin{equation}\label{eq
		}
		\beta(\mathbf{u}) = \arg \min_i \|\mathbf{u} - \mathbf{c}_i\|, \quad \mathrm{with} \quad i=1,\dots,K,
	\end{equation}
	where $\|\cdot\|$ is a norm. The selection of an appropriate distance metric may have an impact on  clustering \cite{Colanera2023,Kelshaw2025}. K-means finds the optimal set of centroids $\mathbf{c}_k$ by minimizing the inner-cluster variance
    \begin{equation}\label{eq:costfun}
		J(\mathbf{c}_1, \ldots, \mathbf{c}_K) = \frac{1}{M} \sum_{m=1}^M \|\mathbf{u}_m - \mathbf{c}_{\beta(\mathbf{u}_m)}\|^2. 
	\end{equation}

\begin{figure}
\centering
\begin{tikzpicture}[node distance=0.34\textwidth, auto]

    \tikzstyle{block} = [rectangle, draw, fill=white!20, text width=0.3\textwidth, text centered, rounded corners, minimum height=5cm]
    \tikzstyle{line} = [draw,  -{Latex[scale=1.15]}]
    
    \node [block] (block1) {\includegraphics[trim= 1cm 1cm 0.35cm 0.7cm,clip,width=0.9\columnwidth]{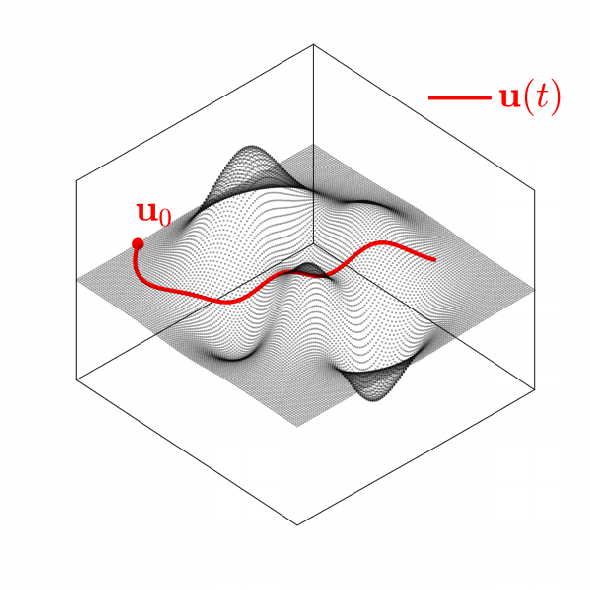}};
    \node [block, right of=block1] (block2) {\includegraphics[trim= 1cm 1cm 0.35cm 0.7cm,clip,width=0.9\columnwidth]{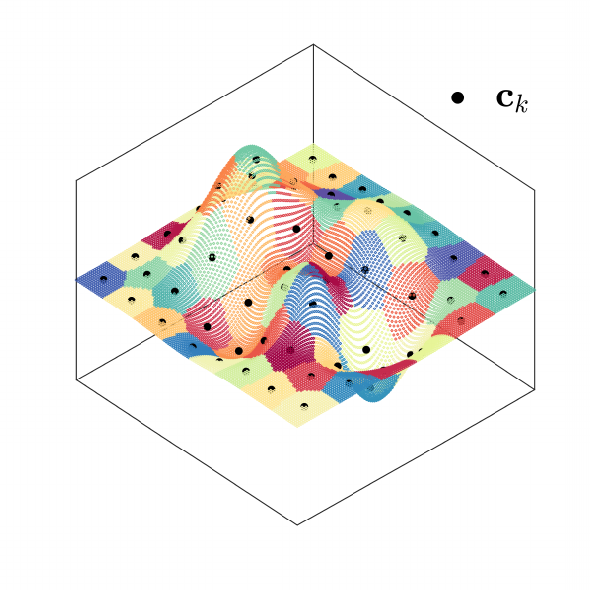}};
    \node [block, right of=block2] (block3) {\includegraphics[trim= 1cm 1cm 0.35cm 0.7cm,clip,width=0.9\columnwidth]{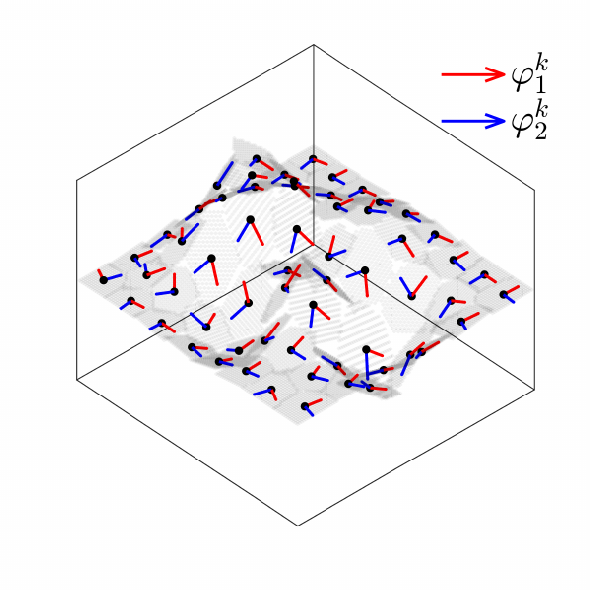}};
    \node [block, below of=block2, node distance=6.3cm,xshift=0.17\textwidth] (block4) {\includegraphics[trim= 0cm 0cm 0cm 0cm,clip,width=0.9\columnwidth]{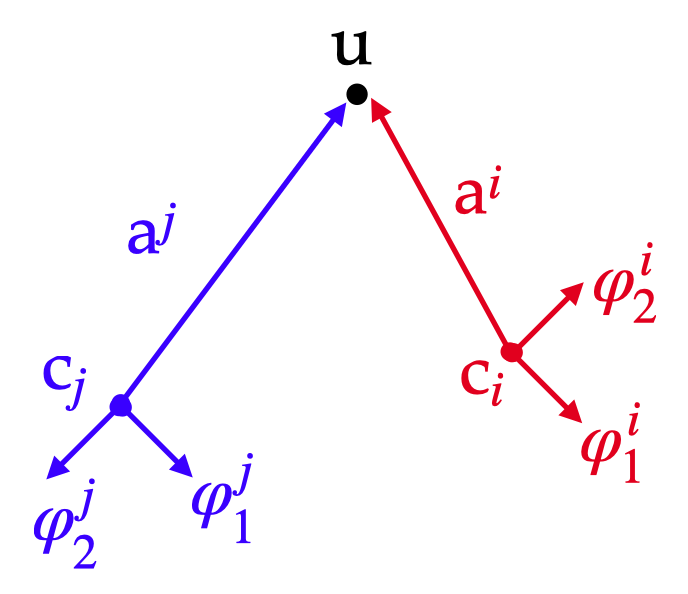}};
 \node [block, below of=block2, node distance=6.3cm,xshift=-0.17\textwidth] (block5) {\includegraphics[trim= 1cm 1cm 0.35cm 0.7cm,clip,width=0.9\columnwidth]{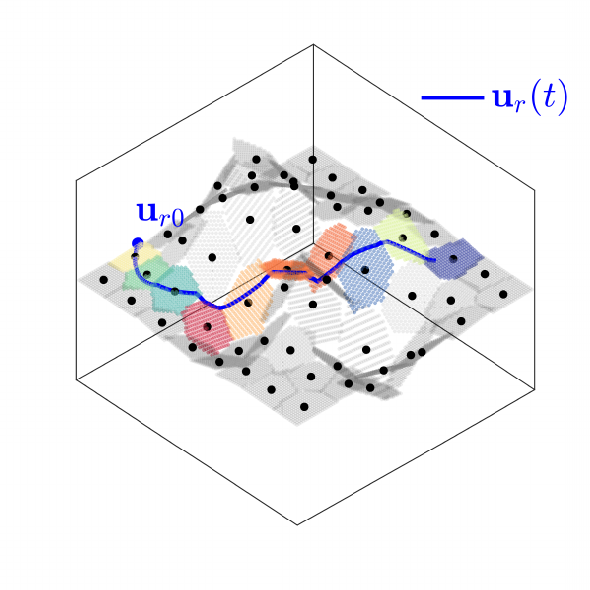}};

 \node [block, above of=block1,minimum height=0.7cm, node distance=2.95cm,text width=0.25\textwidth,xshift=0.025\textwidth] (blockdata) {Data collection};
 \node [block, left of=blockdata,minimum height=0.7cm, node distance=0.16\textwidth,text width=0.03\textwidth] (blockdata) {1};
 \node [block, above of=block2, minimum height=0.7cm, node distance=2.95cm,text width=0.25\textwidth,xshift=0.025\textwidth] (blockdata) {Phase space quantization};
 \node [block, left of=blockdata, minimum height=0.7cm, node distance=0.16\textwidth,text width=0.03\textwidth] (blockdata) {2};
 \node [block, above of=block3, minimum height=0.7cm, node distance=2.95cm,text width=0.25\textwidth,xshift=0.025\textwidth] (blockdata) {Local basis construction};
 \node [block, left of=blockdata, minimum height=0.7cm, node distance=0.16\textwidth,text width=0.03\textwidth] (blockdata) {3};
 \node [block, above of=block4, minimum height=0.7cm, node distance=2.95cm,text width=0.25\textwidth,xshift=0.025\textwidth] (blockdata) {Cluster transition};
\node [block, left of=blockdata, minimum height=0.7cm, node distance=0.16\textwidth,text width=0.03\textwidth] (blockdata) {4.2};
 \node [block, above of=block5, minimum height=0.7cm, node distance=2.95cm,text width=0.25\textwidth,xshift=0.025\textwidth] (blockdata) {Prediction};
\node [block, left of=blockdata, minimum height=0.7cm, node distance=0.16\textwidth,text width=0.03\textwidth] (blockdata) {4.1};

\end{tikzpicture}
\caption{
Schematic overview of the quantized local reduced-order modeling (ql-ROM) framework. The manifold schematically represents the high-dimensional attractor on which the system dynamics evolve. The approach consists of four main stages: (1)~data collection, where trajectories in the state space are sampled; (2)~phase-space quantization, where the manifold is partitioned into clusters; (3)~local basis construction, where a reduced-order model is built around each cluster centroid (illustrated by local 2D patches); and (4)~model deployment, which includes (4.1)~prediction using the active local ROM and (4.2)~cluster transition through coordinate transformations between local bases.
}\label{fig:localROM}
\end{figure}
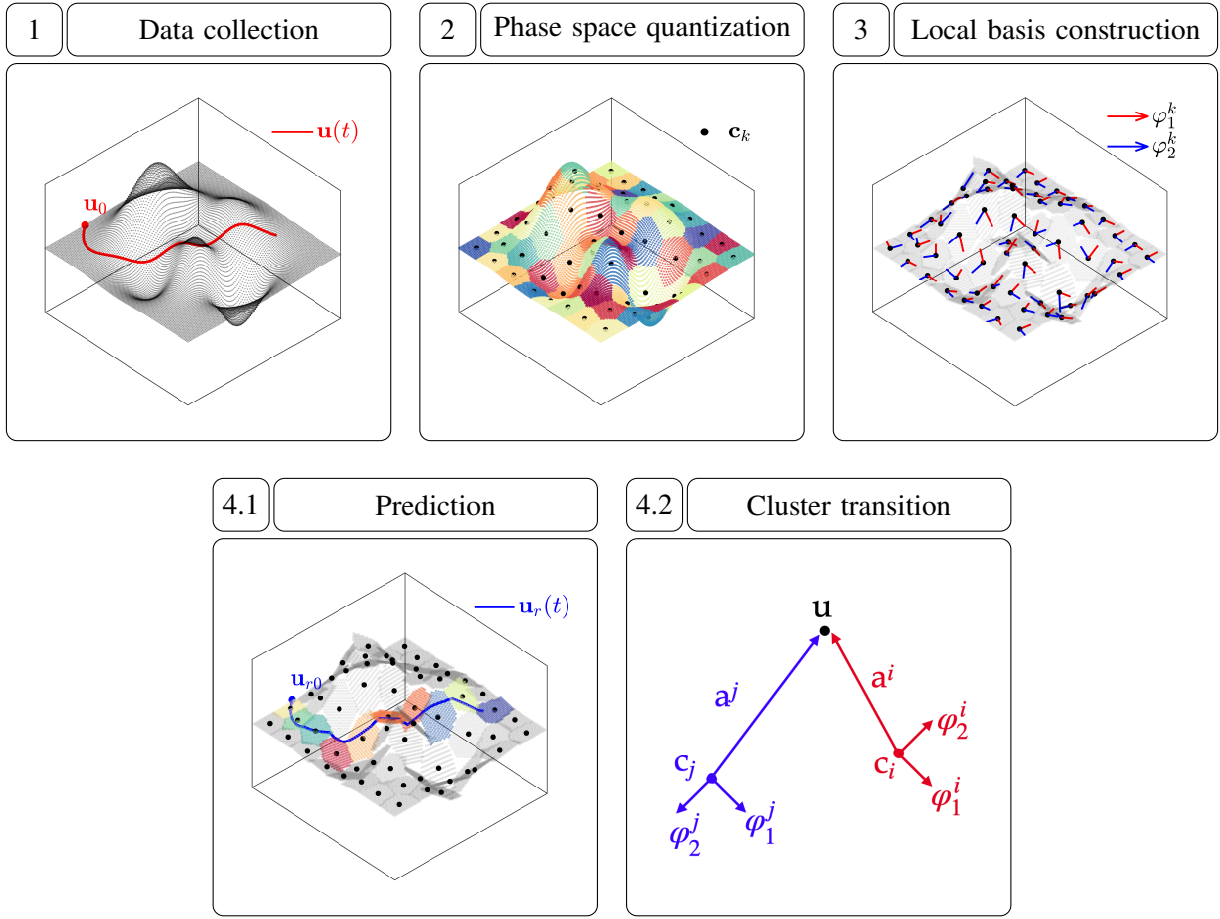

Once the velocity centroids have been computed, the pressure ones ${c_p}_k$ are defined by solving $K$ Poisson problems for each local velocity mean $\mathbf{c}_k$
\begin{equation}
    \nabla^2 {c_p}_k
= -\,\nabla\!\cdot\!\big[(\mathbf{c}_k\!\cdot\!\nabla)\mathbf{c}_k\big].
\end{equation}
For each cluster $k$, we design a local ROM based on the snapshots that belong to that cluster. The local ROM is constructed using the POD snapshot method \cite{Berkooz1993}, which identifies the most energetic modes, in an $L_2$ norm sense, within the cluster. 
The local POD decomposition ansatz reads
\begin{align}
&\mathbf{u}(t) \approx \mathbf{c}_k + \sum_{i=1}^{r_k} a_i^k(t)\,\mathbf{\varphi}_i^k,
\qquad k=\beta(\mathbf{u}(t)),\label{eq:local_ansatz_mfuu}\\
&p(t) \approx {c_p}_k + \sum_{i=1}^{{r_p}_k} b_i^k(t)\,\mathbf{\psi}_i^k,
\qquad k=\beta(\mathbf{u}(t)).
\label{eq:local_ansatz_mfu}
\end{align}
Galerkin projection
within cluster $k$, with $k=1,\,\ldots\,K$, yields a local ROMs
\begin{equation}
\frac{\mathrm{d}\mathbf{a}^k}{\mathrm{d}t}
= \mathcal{F}^k(\mathbf{a}^{k}, \mathbf{b}^{k})
\qquad \text{subject to}\quad \mathcal{G}^{k}(\mathbf{a}^{k}, \mathbf{b}^{k})=0\qquad\mathbf{a}^k \in \mathbb{R}^{r_k},\quad \mathbf{b}^k \in \mathbb{R}^{{r_p}_k},
\label{eq:local_rom_mfu}
\end{equation}
where $\mathcal{F}(\cdot)$ and $\mathcal{G}(\cdot)$ are the local counterpart of eq.~\eqref{eq:rom_mfu} and eq.~\eqref{eq:pressure_rom_mfu}, respectively.
The system’s dynamics evolve along a trajectory confined to a low-dimensional attractor in phase space. As the manifold is locally patched, the state transitions between clusters based on the nearest centroid determined by the cluster-affiliation function. When the nearest centroid changes between time steps $t_m$ and $t_{m+1}$, i.e., $ \beta(m+1) \neq \beta(m) $, the model switches from the ql-ROM centered at $ \mathbf{c}_{\beta(m)} $ to that at $ \mathbf{c}_{\beta(m+1)} $. Consequently, a coordinate transformation of the reduced velocity and pressure coefficients is performed to express the state at $t_{m+1}$ in the new cluster basis. For the POD based ql-ROM, the changes of coordinates are
\begin{align}
&\mathbf{a}^{\,j} = \mathbf{U}_j^{H}\mathbf{U}_i\,\mathbf{a}^{\,i}
+ \mathbf{U}_j^{H}(\mathbf{c}_i-\mathbf{c}_j),\\
	&\mathbf{b}^{j} = \mathbf{V}_{j}^H\mathbf{V}_i \mathbf{b}^i+\mathbf{V}_{j}^H({c_p}_i-{c_p}_{j}),
\label{eq:coord_switch_mfu}
\end{align}
where $(\cdot)^H$ denotes the conjugate transpose operator, and $\mathbf{U}_i$, $\mathbf{V}_{i}$ are the velocity and pressure POD modes matrices, respectively, associated with cluster $\beta(\mathbf{u}_m)=i$. $\mathbf{U}_j$ and $\mathbf{V}_{j}$ are the POD modes matrices of cluster $\beta(\mathbf{u}_{m+1})=j$. The matrix multiplications in eq.~\eqref{eq:coord_switch_mfu} are computed offline and stored.
The reduced-order model prediction $\mathbf{u}_r(t)$ is obtained by integrating only the ql-ROM associated with the current cluster, identified by the cluster-affiliation function. The model initialization requires an initial condition $\mathbf{u}_{r_0}$, which is projected onto the reduced basis of the nearest cluster as $\mathbf{a}_0^{k_0} = \mathbf{U}_{k_0}^H(\mathbf{u}_{r0}-\mathbf{c}_{k_0})$, where $k_0$ denotes the index of the centroid closest to the initial state. An analogous projection is performed for the pressure field. In this work, the initial condition used for the integration corresponds to the last snapshot of the training dataset. After integrating the reduced coordinates and recording the cluster-affiliation sequence, the full velocity and pressure fields are reconstructed using eq.~\eqref{eq:local_ansatz_mfuu} and eq.~\eqref{eq:local_ansatz_mfu}. For $K=1$, the method reduces to the global POD–Galerkin ROM. A summary of the ql-ROM methodology is shown in Figure~\ref{fig:localROM}.

\subsection{Local modal energy budget approach}
\label{sec:methodology_mfu:budget}
To analyze the energy exchanges among the reduced modes and identify the physical mechanisms driving energy transfer, it is convenient to adopt an energy budget formulation. For the global ROM~\eqref{eq:rom_mfu}, the modal energy is defined as \(E_i = \tfrac{1}{2} a_i^2\).  
By multiplying equation~\eqref{eq:rom_mfu} by \(a_i\), its evolution equation reads~\cite{NOACK2005}:
\begin{equation}
\dot{E}_i
= \underbrace{a_i f_i}_{\text{forcing}}
+ \underbrace{\sum_{j} a_i L_{ij} a_j}_{\text{linear production}}
+ \underbrace{\sum_{j,k} a_i C_{ijk} a_j a_k}_{\text{nonlinear transfer}}
- \underbrace{\nu \sum_j a_i B_{ij} a_j}_{\text{viscous dissipation}}
+ \underbrace{a_i \sum_{j=1}^{r_p} P_{ij} b_j}_{\text{pressure forcing}}
+ \underbrace{a_i \tau_i}_{\text{closure}}.
\label{eq:energy_budget_global}
\end{equation}

Here, the matrix $ B_{ij}$ is symmetric positive semi-definite.  
The nonlinear transfer term satisfies
\(\sum_i \sum_{j,k} a_i C_{ijk} a_j a_k = 0\) for energy-preserving tensors \(C_{ijk}\), i.e., it only redistributes energy among modes without net creation or destruction.  
The linear term can be decomposed into symmetric and antisymmetric parts of \(L_{ij}\), separating pure production from rotational (skew-symmetric) effects.

For the local ROM eq.~\eqref{eq:local_rom_mfu} within cluster \(k\), with reduced coordinates \(\mathbf{a}^k\) and \(\mathbf{b}^k\), the modal energy balance becomes
\begin{equation}
\dot{E}_i^{\,k}
= a_i^k f_i^{\,k}
+ \sum_{j} a_i^k L_{ij}^{\,k} a_j^k
+ \sum_{j,\ell} a_i^k C_{ij\ell}^{\,k} a_j^k a_\ell^k
- \nu \sum_j a_i^k B_{ij}^{\,k} a_j^k
+ a_i^k \sum_{j} P_{ij}^{\,k} b_j^k
+ a_i^k \tau_i^{\,k},
\qquad
E_i^{\,k} = \tfrac{1}{2} (a_i^k)^2.
\label{eq:energy_budget_local}
\end{equation}

The instantaneous intermodal transfer to mode \(i\) from advecting or advected modes \(j,\ell\) is defined as
\[
T_{ij\ell}^{\,k}(\mathbf{a}^k) = a_i^k C_{ij\ell}^{\,k} a_j^k a_\ell^k,
\]
and the viscous dissipation is given by
\[
D_i^{\,k}(\mathbf{a}^k) = \nu \sum_j a_i^k B_{ij}^{\,k} a_j^k.
\]

Cluster-conditioned statistics are obtained by averaging over time intervals where \(\beta(\mathbf{u}(t)) = k\):
\begin{align}
\big\langle \dot{E}_i \big\rangle_k
&= \big\langle a_i f_i^{\,k} \big\rangle_k
+ \sum_j \big\langle a_i^k L_{ij}^{\,k} a_j^k \big\rangle_k
+ \sum_{j,\ell} \big\langle T_{ij\ell}^{\,k} \big\rangle_k
- \big\langle D_i^{\,k} \big\rangle_k
+ \big\langle a_i^k \tau_i^{\,k} \big\rangle_k,
\label{eq:cluster_avg_budget}
\end{align}
where \(\langle \cdot \rangle_k\) denotes the conditional average over cluster \(k\).
In this work, we focus on the intermodal energy transfer
\begin{equation}
\big\langle T_{ij\ell}^{\,k} \big\rangle_k
= C_{ij\ell}^{\,k} \big\langle a_i^k a_j^k a_\ell^k \big\rangle_k,
\end{equation}
and the local modal viscous dissipation
\begin{equation}\label{eq:modaldisip}
\big\langle D_i^{\,k} \big\rangle_k
= \big\langle \nu \sum_j a_i^k B_{ij}^{\,k} a_j^k \big\rangle_k
= \nu\, B_{ii}^{\,k}\, \sigma_i^2,
\qquad \mathrm{with}\qquad
\sigma_i^2 = \big\langle (a_i^k)^2 \big\rangle_k.
\end{equation}

\section{Results}\label{sec:results}
\begin{figure}
    \centering
\subfloat[\phantom{a}\hfill\phantom{a}]{%
    \includegraphics[trim= 0cm 0cm 0cm 0cm,clip,width=0.7\columnwidth]{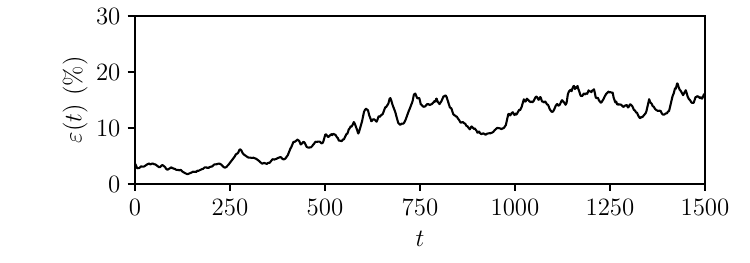}
     }\\
\subfloat[\phantom{a}\hfill\phantom{a}]{%
    \includegraphics[trim= 0cm 0cm 0cm 0cm,clip,width=0.45\columnwidth]{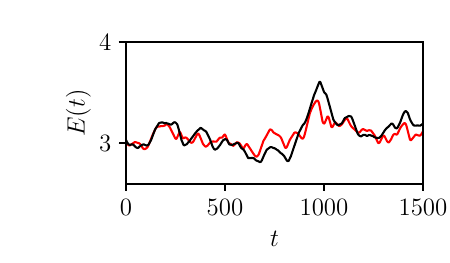}
     }
\subfloat[\phantom{a}\hfill\phantom{a}]{%
    \includegraphics[trim= 0cm 0cm 0cm 0cm,clip,width=0.55\columnwidth]{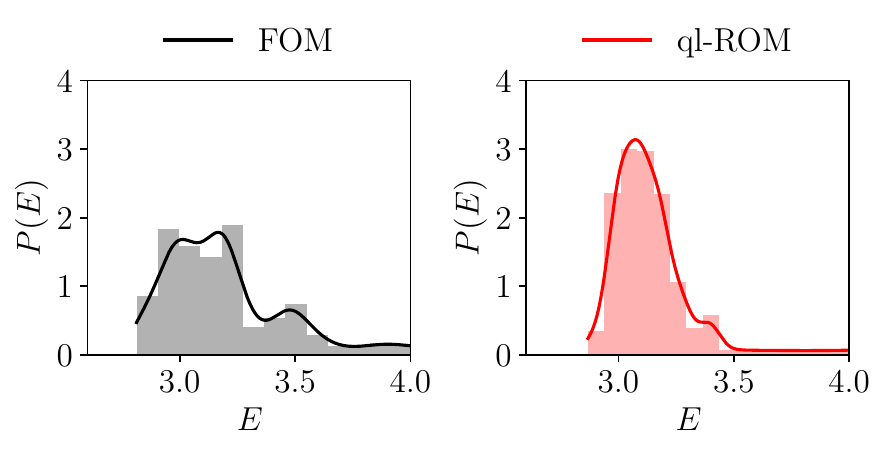}
     }\\
\subfloat[\phantom{a}\hfill\phantom{a}]{%
    \includegraphics[trim= 0cm 0cm 0cm 0cm,clip,width=0.45\columnwidth]{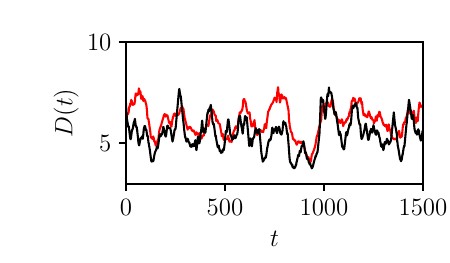}
     }
\subfloat[\phantom{a}\hfill\phantom{a}]{%
    \includegraphics[trim= 0cm 0cm 0cm 0cm,clip,width=0.55\columnwidth]{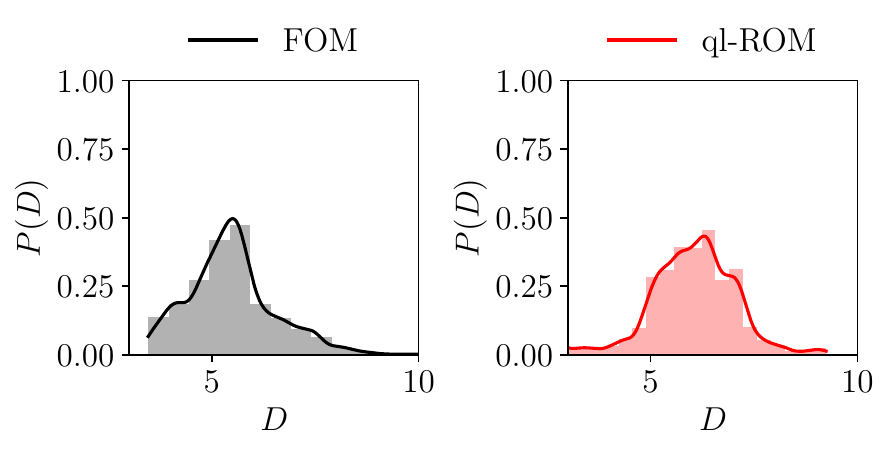}
     }\\
     
    \caption{Performance of the ql-ROM for the Minimal Flow Unit (MFU).
    (a) Prediction error $\varepsilon(t)$, eq.~\eqref{eq:error}, for the test dataset.
    (b) Temporal evolution and PDFs of the total kinetic energy $E(t)$ for the FOM and the ql-ROM.
    (c) PDFs of $E(t)$ for FOM and ql-ROM.
    (d) Temporal evolution of the total dissipation $D(t)$, eq.~\eqref{eq:dissipU},  for both models.
    (e) Corresponding PDFs of $D(t)$ for FOM and ql-ROM.
    The PDFs distributions are estimated using kernel density estimation (KDE) \cite{Scott1992}.
    The ql-ROM exhibits stable and accurate predictions, preserving the main energetic features of the FOM.}
    \label{fig:qlromMFU}
\end{figure}

\begin{figure}
    \centering
\subfloat[\phantom{a}\hfill\phantom{a}]{%
    \includegraphics[decodearray={1 0.  1 0  1 0},trim= 0cm 0cm 10cm 20cm,clip,width=0.57\columnwidth]{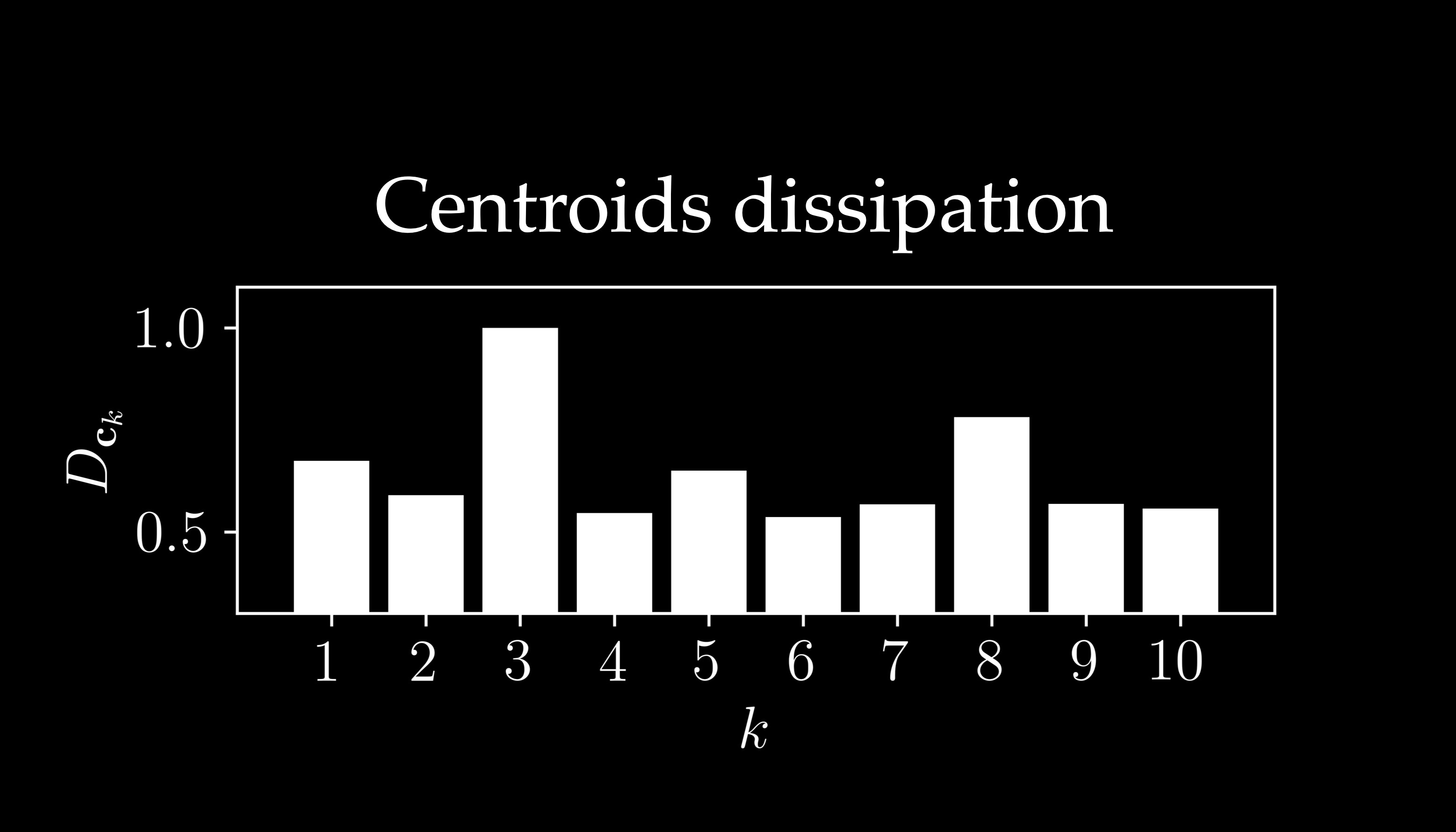}
     }
\subfloat[\phantom{a}\hfill\phantom{a}]{%
    \includegraphics[decodearray={1 0.  1 0  1 0},trim= 0cm 0cm 0cm 17cm,clip,width=0.43\columnwidth]{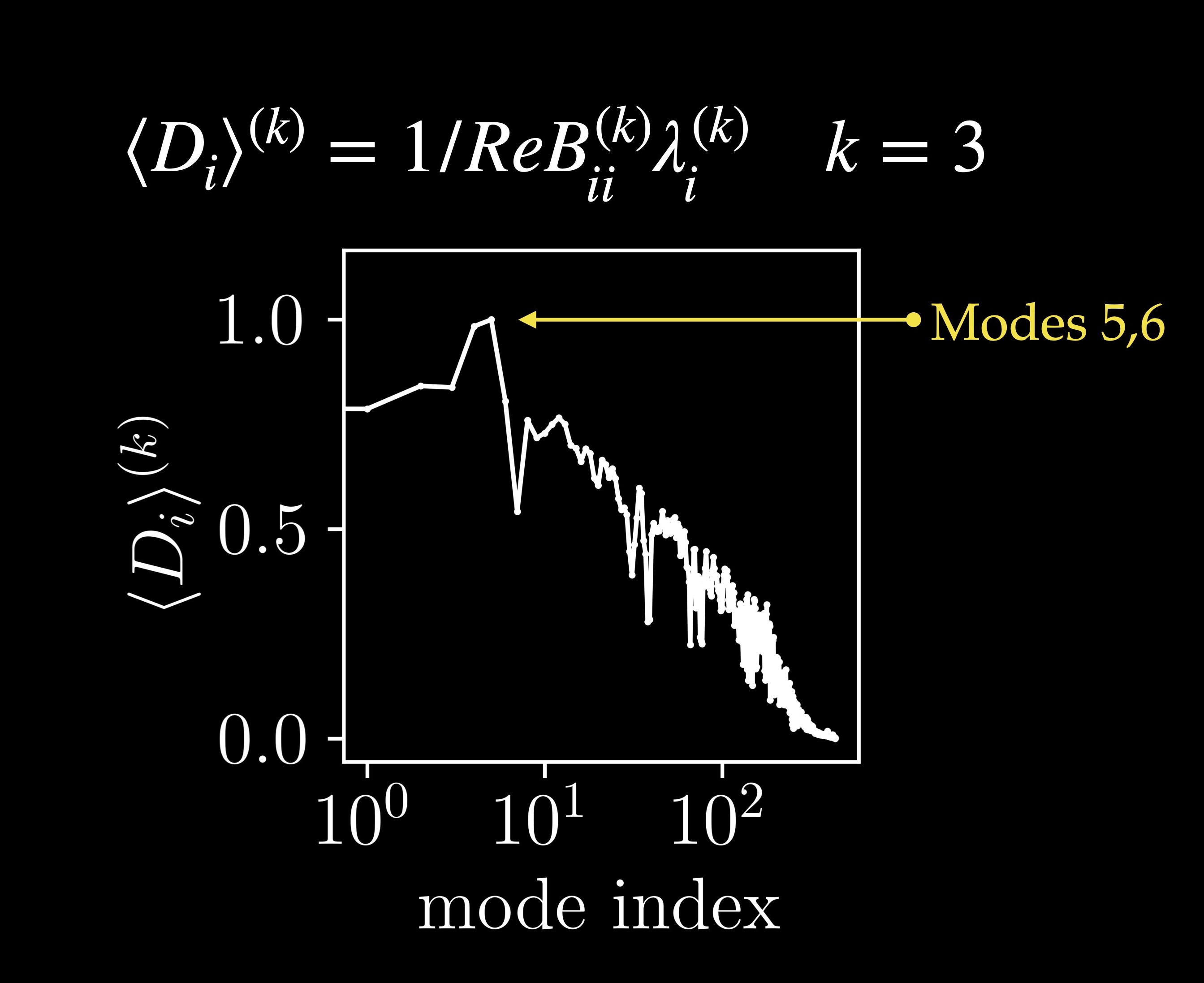}
     }\\
\subfloat[\phantom{a}\hfill\phantom{a}]{%
    \includegraphics[decodearray={1 0.  1 0  1 0},trim= 0cm 0cm 0cm 0cm,clip,width=0.8\columnwidth]{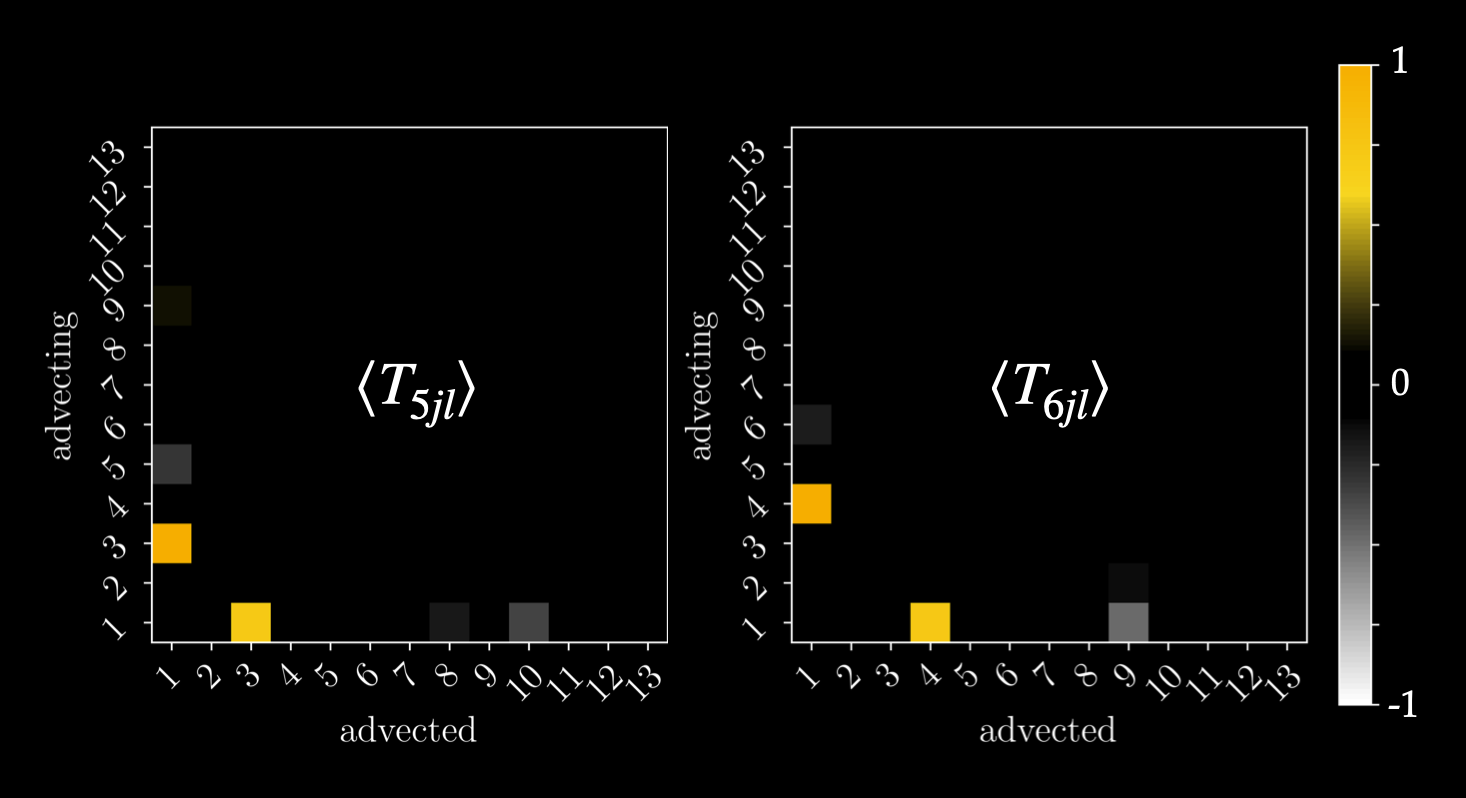}
     }\\
\subfloat[\phantom{a}\hfill\phantom{a}]{%
\includegraphics[decodearray={1 0.  1 0  1 0},trim= 3cm 0cm 0cm 2cm,clip,width=1\columnwidth]{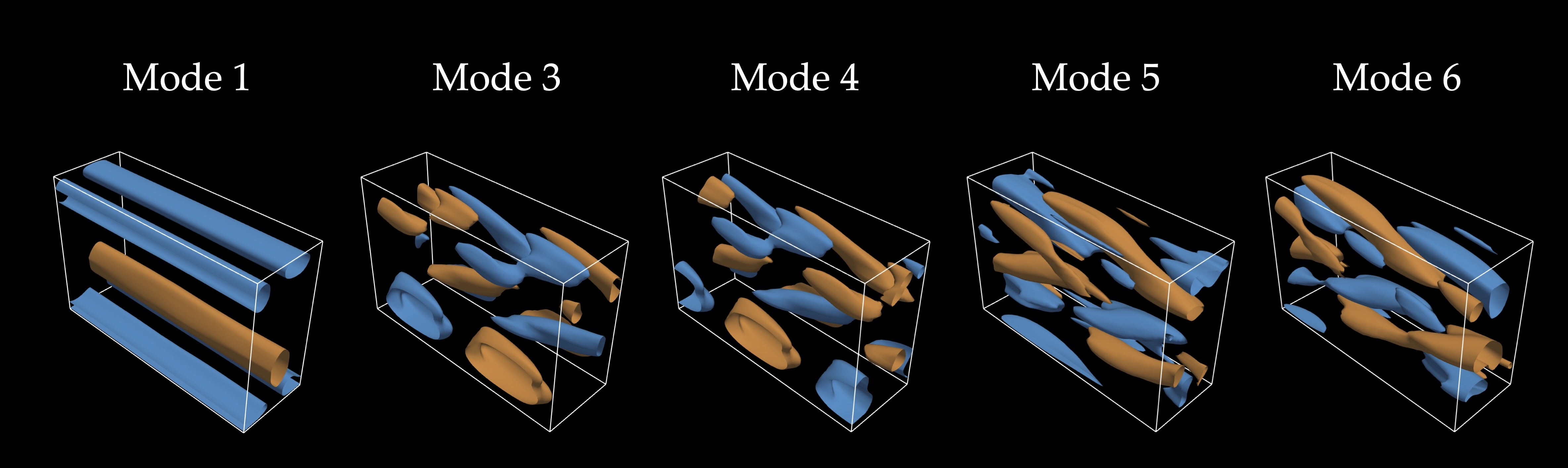}
     }
     
    \caption{  Local modal energy budget analysis of the MFU.  (a) Cluster-averaged dissipation normalized by its maximum value; cluster $k=3$ shows the highest dissipation. 
    (b) Modal dissipation $\langle D_i^{\,k} \rangle_k$ for $k=3$, which shows that modes~5 and~6 are the most dissipative. 
    (c) Selected intermodal energy transfer terms showing the main nonlinear couplings. 
    (d) Axial velocity structures of selected modes.}
    \label{fig:mfuEB}
\end{figure}

In this section, the methodologies introduced in Section~\ref{sec:methods} are applied to the flow field of the Minimal Flow Unit (MFU). 
The training dataset consists of $40{,}000$ snapshots of velocity and pressure fields, sampled with a time step of $\Delta t = 0.2$. 
The test dataset includes $7{,}500$ future snapshots. 
The first step of the analysis involves the construction of a ql-ROM capable of providing accurate and stable predictions. 
Two key hyperparameters of the ql-ROM are the number of clusters and the number of modes per cluster. 
In this work, we selected $K = 10$ clusters, while the number of modes was determined such that the root mean squared reconstruction error at the $m$-th time instance, defined as
\begin{equation}\label{eq:residuo}
	\mathbf{r}_m = (\mathbf{I} - \mathbf{U}_k\mathbf{U}_k^T)(\mathbf{u}_m - \mathbf{c}_k),
	\qquad \mathrm{with} \quad k = \beta(\mathbf{u}_m),
\end{equation}
with $\mathbf{I} \in \mathbb{R}^{N\times N}$ denoting the identity matrix, is below $0.1\%$. 
Accordingly, $r = 500$ modes were retained for each cluster, for both velocity and pressure fields, using the same $r$ across clusters for simplicity.
Figure~\ref{fig:qlromMFU} summarizes the performance of the ql-ROM. 
Panel~(a) shows the prediction error for the test dataset,
\begin{equation}\label{eq:error}
	\varepsilon(t) = \frac{\| \mathbf{u}(t) - \mathbf{u}_r(t) \|}{\| \mathbf{u}(t) \|},
\end{equation}
showing that the model remains stable and the relative error stays below $20\%$ over the considered time window. 
Panel~(b) compares the temporal evolution of the total kinetic energy $E$ in the test dataset for both the FOM and the ql-ROM. 
The energy remains bounded, and the probability density functions (PDFs) of both models exhibit similar mean values, although the ql-ROM distribution tends to be more unimodal compared to the FOM. The PDFs distributions are estimated using kernel density estimation (KDE) \cite{Scott1992}.
Panels~(c) and~(d) display the temporal evolution of the total dissipation and the corresponding PDFs for both models, confirming the good performance and physical consistency of the ql-ROM.
Once the ql-ROM has been designed it is possible to apply the local modal energy budget approach introduced in section \ref{sec:methodology_mfu:budget}. To investigate the transfer mechanism during the dissipation bursts it is convenient to identify which region of the phase space is characterized by high dissipation. For this we have computed the centroids dissipation as in eq.~\eqref{eq:dissipU}
\begin{equation}\label{eq:dissipcentr}
   D_{\mathbf{c}_k} = \frac{\nu}{|\Omega|} \int_{\Omega} |\nabla \times \mathbf{c}_k|^2\, d\Omega.
\end{equation}
Panel~(a) of Figure~\ref{fig:mfuEB} shows the centroid-averaged dissipation of the $K$ clusters, normalized by its maximum value. 
Cluster $k=3$ exhibits the highest dissipation among its representative snapshots, and the analysis will therefore focus on this region of the phase space. 
Within this cluster, it is insightful to identify the modes associated with the largest expected modal dissipation, defined in eq.~\eqref{eq:modaldisip}. 
Panel~(b) shows the values of $\big\langle D_i^{\,k} \big\rangle_k$ for $k=3$, indicating that modes~5 and~6 are the most dissipative in this region of the phase space. 
To further investigate the nonlinear mechanisms responsible for energy redistribution, Panel~(c) of Figure~\ref{fig:mfuEB} presents the intermodal energy transfer terms $\big\langle T_{5j\ell}^{\,k} \big\rangle_k$ and $\big\langle T_{6j\ell}^{\,k} \big\rangle_k$ for the same cluster. 
These matrices highlight the advecting and advected mode pairs that transfer energy to modes~5 and~6 through nonlinear interactions. 
The results reveal that mode~5 primarily receives energy from the interaction between modes~1 and~3, while mode~6 is energized mainly by the coupling between modes~1 and~4. 
Panel~(d) shows the spatial distribution of the axial velocity component for modes~1,~3,~4,~5, and~6. 
Mode~1 corresponds to streak-like structures near the lower wall, modes~3 and~4 are associated with longitudinal travelling-wave motions in the streak region, and modes~5 and~6 display chaotic, highly dissipative vortical structures.

With this analysis the dissipation bursts observed in the MFU can be directly linked to the nonlinear interaction between the streamwise streaks and the travelling-wave modes similarly to what was found in Ref. \cite{Bae2021,Gayme2019}. The nonlinear interaction between these waves and the distorted streaks regenerates new vortices and induces strong velocity gradients, leading to a sudden increase in the viscous dissipation rate.

\section{Conclusions}\label{sec:conclusions}
A quantized local reduced-order modeling (ql-ROM) framework has been applied to the turbulent Minimal Flow Unit (MFU) to predict and interpret intermittent dissipative dynamics (extreme events). The ql-ROM combines clustering-based manifold partitioning with local intrusive Galerkin projection, thereby accounting for the nonlinear and multi-regime nature of turbulent flows while retaining a fully interpretable reduced dynamical structure. The model provides accurate and stable temporal predictions and reproduces the statistical behavior of kinetic energy and dissipation observed in the full-order simulations.

The local modal energy-budget formulation introduced in this work enables a direct quantification of intermodal energy transfers and modal viscous dissipation within each cluster. This analysis has shown that high-dissipation regions of the phase space are characterized by intensified nonlinear energy transfer toward a small subset of modes associated with strongly vortical structures. In the MFU, these modes correspond to the interaction between streamwise streaks and travelling-wave motions, confirming their key role in the regeneration cycle and in the onset of dissipation bursts.

Beyond model reduction and interpretability, the ql-ROM framework provides a systematic and computationally efficient strategy for identifying and analyzing extreme events in turbulence. The partition of the attractor into dynamically coherent regions allows the association of specific clusters with high-dissipation states, while transitions between clusters can serve as precursors of forthcoming bursts. The conditional statistics of modal dissipation and nonlinear transfer further enable the identification of energy pathways leading to extreme behavior, offering a reduced-space representation of the mechanisms underlying intermittent events.

The results demonstrate that the ql-ROM is a predictive and diagnostic tool for the study of extreme events in wall-bounded turbulence. Future developments will focus on the quantitative assessment of its predictive capability through probabilistic/deterministic forecasting of dissipation bursts.

\section*{Acknowledgments}
This work was supported in part by the European Research Council under the Caust grant ERC-AdG-101018287. We acknowledge also the support from the grant EU-PNRR YoungResearcher TWIN ERC-PI\_0000005.

\bibliography{references} 
\end{document}